

\font\mybb=msbm10 at 12pt
\def\bb#1{\hbox{\mybb#1}}
\def\Z {\bb{Z}}
\def\R {\bb{R}}

\tolerance=10000
\input phyzzx

 \def\unit{\hbox to 3.3pt{\hskip1.3pt \vrule height 7pt width .4pt \hskip.7pt
\vrule height 7.85pt width .4pt \kern-2.4pt
\hrulefill \kern-3pt
\raise 4pt\hbox{\char'40}}}

\def\SS{{{\cal S}}}
\def\HH{{\cal H}}

\def\gmn{{g _{\mu \nu}}}

\def\b {{\psi}}

\REF\HT{C.M. Hull and P.K. Townsend, Nucl. Phys. {\bf B438} (1995) 109.}
\REF\PKT{P.K. Townsend, Phys. Lett. {\bf B350}, (1995) 184.}
\REF\Witten{E. Witten, {\it String Theory Dynamics in Various Dimensions},
hep-th/9503124.}
\REF\HTE{C.M. Hull and P.K. Townsend, {\it Enhanced Gauge Symmetries in
Superstring Theories}  (revised), hep-th/9505073.}
\REF\Seno{A. Sen, {\it String-String Duality Conjecture in Six Dimensions and
Charged Strings}, hep-th/9504027.}
\REF\JHS {  J.A.  Harvey and A. Strominger,  {\it The Heterotic String is a
Soliton}, hep-th/9504047.}
\REF\duff{M.J. Duff, {Strong/Weak Coupling Duality from the Dual String},
hep-th/9501030.}
\REF\FILQ{A. Font, L. Ibanez, D. Lust and F. Quevedo, Phys. Lett. {\bf B249}
(1990) 35; S.J. Rey, Phys. Rev. {\bf D43}  (1991) 526; J.H. Schwarz and A. Sen,
Nucl. Phys. {\bf B411} (1994) 35; Phys.
Lett. {\bf 312B} (1993) 105;  A. Sen,   Int. J. Mod. Phys. {\bf A9} (1994)
3707.}
\REF\DGHR {A. Dabholkar, G.W. Gibbons, J.A. Harvey and F. Ruiz-Ruiz,
Nucl. Phys. {\bf B340} (1990) 33.}
\REF\HS {G.T. Horowitz and A. Strominger, Nucl. Phys. {\bf B360} (1991)
197.}
\REF\GT {G.W. Gibbons  and P.K. Townsend, Phys. Rev.  Lett.
{\bf  7 }  (1993) 3754.}
\REF\DGT {M.J. Duff, G.W. Gibbons and P.K. Townsend, Phys. Lett. {\bf 332 B}
(1994) 321.}
\REF\GHT {G.W. Gibbons, G.T. Horowitz and P.K. Townsend, Class. Quantum Grav.
{\bf 12} (1995) 297.}
\REF\paul{P.K. Townsend, private communication.}
\REF\CHS {C. Callan, J. Harvey and A. Strominger, Nucl. Phys. {\bf B359}
(1991)
611.}
\REF\DL {M.J. Duff and J.X. Lu, Nucl. Phys. {\bf B354} (1991) 141;
Phys. Lett. {\bf 273B} (1991) 409.}
\REF\bergort{E. Bergshoeff, C.M.Hull and T. Ortin, {\it Duality in the Type II
Superstring Effective Action}, hep-th/9504081.}
\REF\Sentrans{H. Hassan and A. Sen, Nucl. Phys. {\bf B375} (1992) 103,
hep-th/9109038;
H. Hassan, hep-th/9408060}.
\REF\Smacrto{A. Sen, Nucl. Phys. {\bf B388} (1992) 457.}
\REF\KKhet{S.J. Gates and W. Siegel,  Phys. Lett. {\bf 206B}  (1988) 631;  S.J.
Gates, S.V. Ketov, S.M. Kuzenko and O.A. Soloviev,
Nucl. Phys. {\bf B362} (1991) 199.}
\REF\Strom{A. Strominger,  Nucl. Phys. {\bf B343} (1990) 167.}
\REF\dab{A. Dabholkar, {\it Ten-Dimensional Heterotic String as a Soliton},
hep-th/9506160.}


\Pubnum{ \vbox{ \hbox {QMW-95-25}  \hbox{hep-th/9506194}} }
\pubtype{}
\date{June 1995}

\titlepage

\title {\bf 	STRING-STRING DUALITY IN TEN DIMENSIONS}

\author{C.M. Hull}
\address{Physics Department,
Queen Mary and Westfield College,
\break
Mile End Road, London E1 4NS, U.K.}

\abstract{The heterotic string occurs as a soliton of the type I superstring in
ten dimensions, supporting the conjecture that  these two theories are
equivalent.
The conjecture that the type IIB string is self-dual, with the strong coupling
dynamics described by a dual  type IIB theory, is supported by the occurrence
of the dual string as a Ramond-Ramond soliton of the weakly-coupled theory.
}

\endpage
\pagenumber=1

\def\so{{$SO(32)$}}
\def\af{{1- \left(a \over r \right) ^6}}
\def\afb{{\left( \af \right)}}

\chapter{Introduction}

Recent evidence suggests that, in many cases, string theories that appear very
different perturbatively may in
fact be equivalent
when non-perturbative effects are taken into account [\HT-\HTE].
For example, the heterotic string compactified to six dimensions on $T^4$ is
conjectured to be equivalent to the type II string compactified on $K_3$ [\HT].
Soon after the heterotic string was discovered, it was conjectured by   Green
and Witten that the
heterotic string with gauge group
$SO(32)$ and the type I string with the same gauge group might be equivalent;
indeed, they have the
same low-energy effective field theory. The heterotic string spectrum contains
spinor
representations of  \so, but there are no such representations in the
perturbative spectrum of the
type I string, so such representations would have to arise as solitons of the
type I theory. It was
pointed out in [\Witten] that if such a relation were to hold, then the strong
coupling limit of one
should correspond to the weak coupling limit of the other. This is certainly
the case at the level
of the low-energy effective field theories [\Witten]. In this paper further
evidence supporting this
conjecture, and  the conjectured self-duality of type
IIB strings [\HT,\Witten], will be provided.

Evidence supporting the conjectured equivalence of type II and heterotic
strings in six dimensions was presented in  [\Seno,\JHS],  where it was shown
that the heterotic string emerges as a non-singular soliton of the type II
string compactified on $K_3$ with the correct zero-mode structure. It is
believed that the type II string should also arise as a  soliton of the
six-dimensional heterotic string, but the zero-mode analysis of the natural
solution in this context has so far proved problematic [\JHS].
Such  situations in which the strong coupling limit of one string theory is the
weak coupling limit of another \lq dual'  string theory   were discussed
earlier by Duff [\duff], who called the phenomenon string-string duality.
In the same spirit, it will be shown here that the ten-dimensional heterotic
string arises as a   soliton of the type I string, which is   evidence in
favour of the conjectured heterotic/type I equivalence.  To complete the
picture, it would be desirable  to obtain the type I string as a
soliton of the heterotic string theory, but as yet no such soliton is known.

The heterotic string in four dimensions is believed to be self-dual: the strong
coupling limit is described by a dual heterotic string with electric and
magnetic charges interchanged, generalising to string theory the conjectured
Montonen-Olive self-duality of $N=4$ supersymmetric Yang-Mills.  This is a
consequence of the conjectured $SL(2,\Z)$ S-duality of the theory [\FILQ]. In
[\HT], it was conjectured that the type IIB superstring in ten dimensions also
has an
$SL(2,\Z)$ duality symmetry. This would imply that this theory is also
self-dual, as the $SL(2,\Z)$ symmetry includes a transformation that
interchanges weak and strong coupling regimes. However, whereas the
four-dimensional heterotic string duality interchanges electric and magnetic
charges, this ten-dimensional duality interchanges Neveu-Schwarz/Neveu-Schwarz
(NS-NS) and Ramond-Ramond (RR) charges.
It will be shown here that the weakly coupled type IIB string has a   solitonic
string that is interchanged with
the fundamental type IIB string by duality and which corresponds to the
fundamental string of the dual strongly coupled theory.
This constitutes further evidence in favour of the conjectured self-duality of
the type IIB string.

\chapter {Solitons of Heterotic and Type I Strings}

The heterotic string low-energy effective action
includes the bosonic terms
$$
\int
 d^{10} x \sqrt{-g} e^{-2 \Phi} \left( R+ 4\partial _ \mu \Phi \partial ^ \mu
\Phi -{1 \over 4}
F_{\mu \nu}^I
 F^{I \mu \nu }- {1 \over 12}H_{ \mu \nu \rho }H^{ \mu \nu \rho }
\right)
\eqn\het$$
while the corresponding action for the type I string is [\Witten]
$$
\int
 d^{10} x \sqrt{-g}  \left(e^{-2 \Phi } \left( R+ 4\partial _ \mu \Phi
\partial ^ \mu \Phi
\right)-{1 \over 4} e^{- \Phi }F_{\mu \nu}^I
 F^{I \mu \nu }- {1 \over 12}H_{ \mu \nu \rho }H^{ \mu \nu \rho }
\right)
\eqn\open$$
 with different dilaton dependence. The substitution
$$ g_{\mu \nu}\to e^\Phi g_{\mu \nu}, \qquad \Phi \to- \Phi
\eqn\subs$$
in the heterotic action \het\ gives the type I action \open. The dilaton sign
flip implies that the
heterotic string coupling constant    is inversely related to the type I string
coupling constant,
 so that the weak coupling limit of one indeed corresponds to the strong
coupling
limit of the other [\Witten].

The heterotic string has a neutral string soliton
given by [\DGHR]
$$\eqalign{
ds ^2&= \afb (-dt^2 + d \sigma ^2) +\left( \af \right) ^{-5/3} dr^2 + r^2 \afb
^{1/3}
d \Omega _7^2
\cr
e^{2\Phi}&= \af, \qquad
H=6 a^6 e^{2\Phi} *\epsilon _7
\cr}
\eqn\strsol$$
where   $\sigma$ is the coordinate parameterising
 the string, $*$ denotes the Hodge dual,
$d \Omega _7^2$ is the line element on the unit seven-sphere
and
$\epsilon _7$ is its volume form.
This is the extremal limit of the black string solutions of [\HS] and is
singular on the horizon
$r=a$. It also preserves half of the ten-dimensional supersymmetry and
saturates the corresponding
Bogomolnyi bound [\DGHR]. The solution describes the space-time outside
a fundamental string source [\DGHR] and might be thought of as a soliton which
should be identified with
the fundamental string.

There is also a neutral string solution of the type I string action
\open\ given by
$$\eqalign{
d{s} ^2&=
A ^{1/2} (-dt^2 + d \sigma ^2) + A ^{-1/6 }\left( A^{-2} dr^2 + r^2   d
\Omega _7^2 \right)
\cr
e^{-2\Phi }&= A, \qquad A \equiv \af
, \qquad
H=6 a^6 e^{-2\Phi } *\epsilon _7
\cr}
\eqn\opsol$$
This is related to \strsol\ by the conformal rescaling  of the metric and
dilaton sign flip \subs.
The metric is the special case $\alpha =-1/2$ of the   rescaled metric
$$\tilde g _{\mu \nu} =e^{2\alpha \Phi }g _{\mu \nu}
\eqn\gti$$
and the singularity or otherwise of this class of metrics will now be
considered.

The metric
$$
d\tilde s ^2= A ^{\alpha+1} (-dt^2 + d \sigma ^2) + A ^{\alpha+1/3 }\left(
A^{-2} dr^2 + r^2   d
\Omega _7^2 \right)
 \eqn\gensol$$
(with $A$ given by \opsol)
has a potential singularity at $r=a$. To investigate the behaviour near $r=a$,
it is convenient
to follow [\GT-\GHT] and define a new coordinate
$$\lambda = {6 (r-a) \over a}
\eqn\lamis$$
so that the metric becomes
$$
d\tilde s ^2= \lambda ^{\alpha+1} (-dt^2 + d \sigma ^2)[1+O(\lambda)] +
\lambda ^{\alpha+1/3 }\left(
{a^2 \over 36}\lambda^{-2} d\lambda ^2 + a^2   d
\Omega _7^2 \right) [1+O(\lambda)]
\eqn\gensola$$
Further defining
$$
\rho = {1\over 6} a \ln \lambda
\eqn\rhis$$
and suppressing the $O(\lambda)$ terms, the metric tends asymptotically to
$$
d\tilde s ^2 \sim  e^{ {6\over a}(\alpha+1) \rho}    (-dt^2 + d \sigma
^2) +
e^{ {6\over a}(\alpha+1/3) \rho}
 \left(
  d\rho ^2 + a^2   d
\Omega _7^2 \right)
\eqn\gensolb$$
as $r \to a$, $\rho  \to - \infty$.
Asymptotically, the dilaton is given by
$$ \Phi \sim {3 \over a} \rho
\eqn\Phiis$$
so that  there is a linear dilaton (in these coordinates) which blows up as  $r
\to a$, $\rho  \to
- \infty$.
Thus there is a dilaton singularity as well as a potential curvature
singularity at $r=a$, so that
the solution can only be regarded as non-singular if   $r=a$ is an infinite
geodesic distance from
all points with
$r>a$.

There are a number of different cases, depending on the value of $\alpha$.
The asymptotic metric is a warped product of two-dimensional Minkowski-space
$M_2$
with coordinates $t, \sigma$ and a cylindrical \lq throat' $C_8=\R \times S^7$
which has  a
seven-sphere cross-section of radius $a$. If $\alpha <-1$ the conformal factors
blow up both $M_2$
and $C_8$ as $\rho \to - \infty$,  while if $-1 < \alpha < - 1/3$ the throat
opens up (like the bell of a trumpet)
and the
$M_2$ shrinks. If $\alpha > -1/3$ both  $C_8$ and $M_2$ contract, so that the
throat pinches
off. If $\alpha < -1/3$, then  it follows from the asymptotic form of the
metric \gensolb\ that the distance along a space-like geodesic lying entirely
within $C_8$  from any point with finite
$\rho$ to $\rho =- \infty$, which corresponds to $r=a$,  is   infinite. Thus a
solution with  $\alpha < -1/3$, and in particular the type I solution \opsol\
with $\alpha = -1/2$, has
non-singular spatial sections of constant $t, \sigma$.
For $\alpha \ge -1/3$, it is a finite distance to $r=a$ along space-like
geodesics, so that the solution is singular.
In the special case $\alpha = -1/3$, the asymptotic metric is the standard
metric on $S^7 \times
AdS^3$, where $AdS^3$ is three-dimensional anti-de Sitter space [\DGT]. In this
case, $\tilde g _{\mu
\nu}$ is the so-called 5-brane metric  and the string soliton has a
non-singular geometry which
interpolates between $d=10$ Minkowski space $M_{10}$ and the $S^7 \times
AdS^3$ string solution with linear dilaton, but has a dilaton that diverges on
$r=a$ [\DGT].

Although $r=a$ cannot be reached along space-like geodesics of finite length
for $\alpha <-1/3$, there remains the possibility of the singularity being
reached by time-like or null geodesics of finite length.  In fact, the
singularity can be reached by a finite-length time-like   geodesic
if $\alpha > -2/3$ [\paul].
For a time-like radial geodesic $r(t)$, the proper time $\tau$ is defined by
$$
d \tau ^2 = A^{ \alpha +1} dt^2 - A^{ \alpha - 5/3} dr^2
\eqn\dtis$$
and energy conservation implies that
$$
A^{ \alpha +1} {dt  \over d \tau}= \epsilon
\eqn\encon$$
for some constant $ \epsilon$.
This implies
$$ \left( { \epsilon^2 \over  A^{ \alpha +1}} -1 \right) d \tau ^2 = A^{ \alpha
- 5/3} dr^2
\eqn\dtmo$$
Near $r=a, \lambda=0$, $A \sim \lambda$ and if $\alpha >-1$, then \dtmo\
implies
$${ d \tau  \over d \lambda}
\sim {a \over  6 \epsilon} \lambda ^{ \alpha -1/3}
\eqn\dtasym$$
so that the proper time along the geodesic from the singularity $\lambda =0$ to
some small value $ \lambda= \Lambda$ is
$$ \tau \sim {a \over  6 \epsilon} \int
_ 0 ^ \Lambda { d \lambda \over \lambda ^{ 1/3- \alpha}}
\eqn\Intis$$
and this is finite if $ \alpha > -2/3$.
In particular, for the open string soliton \opsol, $ \alpha= -1/2$ and the
singularity can be reached along time-like geodesics.

The heterotic string action \het\ also has a five-brane solution [\CHS,\DL]
$$\eqalign{
ds ^2&=  -dt^2 + dx^i dx^i +B^{-2} dr^2 + r^2
d \Omega _3^2
\cr
e^{-2\Phi}&= B, \qquad B= {1- \left(a \over r \right) ^2}
, \qquad
H=2 a^2  \epsilon _3
\cr}
\eqn\fivsol$$
where $x^i$ ($i=1,  \dots 5$)  are coordinates on the five-brane,
$d \Omega _3^2$ is the line element on the unit three-sphere
and
$\epsilon _3$ is its volume form. This is a stable non-singular soliton that
saturates a Bogomolnyi
bound and interpolates between the Minkowski space $M_{10}$ and the $S^3$
compactification to
$d=7$ Minkowski space with a linear dilaton [\GT]. Consider
the   rescaled metric \gti, which in this case is given by
$$
d\tilde s ^2=B ^{-\alpha  }\left(-dt^2 + dx^i dx^i  +  B^{-2} dr^2 + r^2   d
\Omega _3^2 \right)
\eqn\gensolfiv$$
and  has a potential singularity at $r=a$. To investigate the behaviour near
$r=a$, it is convenient
to again follow [\GT-\GHT] and define   new coordinates
$$\lambda = {2 (r-a) \over a}, \qquad \rho = {1\over 2} a \ln \lambda
\eqn\lamisfi$$
so that  suppressing the $O(\lambda)$ terms, the metric tends asymptotically to
$$
d\tilde s ^2 \sim  e^{ -{2\alpha\over a}  \rho}
\left(  -dt^2 + dx^i dx^i +
d\rho ^2 + a^2   d
\Omega _3^2 \right)
\eqn\gensolbfi$$
as $r \to a$, $\rho  \to - \infty$.
Asymptotically, the dilaton is given by
 $ \Phi \sim -{1 \over a} \rho
 $
so that  there is a linear dilaton   which blows up as  $r \to a$, $\rho  \to
- \infty$.
{}From the asymptotic form
\gensolbfi, the distance to $r=a$   along space-like geodesics  is   finite
if $\alpha <0$. Thus although the five-brane is a non-singular solution of the
heterotic string ($\alpha=0$),
it is   singular in the open string metric ($\alpha =-1/2$) and the five-brane
metric
($\alpha =-1/3$).

\chapter{Solitons of the Type IIB String}

The massless bosonic fields of the type IIB string consist of the graviton
$\bar  g_{\mu \nu }$, the
NS-NS anti-symmetric tensor gauge field $ b_{\mu \nu } ^1$, the RR one $ b
_{\mu \nu }^2$, a four-form   with self-dual field strength $G_{ \mu \nu \rho
\sigma \tau}$ and a complex scalar
$\lambda = \b+ie^{-\Phi}$, where $\Phi$ is the dilaton.
There is no action that  gives the full low-energy field equations, but for
solutions with
  $G_{ \mu \nu \rho
\sigma \tau}=0$, the remaining bosonic low-energy field equations can be
obtained by varying the
action
$$
\int
 d^{10} x \sqrt{-\bar g}   \left( \bar R+ 4 tr \partial _ \mu \SS \partial ^
\mu \SS  ^{-1}-  {1
\over 12}tr \left[ \HH^t_{
\mu \nu \rho }\SS  \,  \HH^{ \mu \nu \rho } \right]
\right)
\eqn\twob$$
where $\bar g_{\mu \nu}$  is the canonical Einstein metric and
$$
{\cal S} = { 1 \over \lambda _2}\pmatrix { | \lambda|^2 & \lambda _1 \cr
\lambda _1 & 1}, \qquad \HH_{ \mu \nu \rho }=\pmatrix {H_{ \mu \nu \rho }^1 \cr
H_{ \mu \nu \rho }^2 }, \qquad \lambda =\lambda _1 + i
\lambda_2, \qquad H^i=db^i
\eqn\sis
$$

The type IIB supergravity field equations are invariant under $SL(2,\R)$ which
leaves $G$
and  $\bar g_{\mu \nu}$ invariant and acts on the remaining fields by
$$
\SS \to  \Lambda \SS \Lambda ^t, \qquad \HH \to \Lambda^{-1} \HH
\eqn\sltran$$
where $\Lambda $ is a $2\times 2$ matrix in  $SL(2,\R)$.
In the string metric $\gmn =e^{\Phi /2}\bar  g_{\mu \nu }$, the action is
[\bergort]
$$
\int
 d^{10} x \sqrt{-g} \left[ e^{-2 \Phi} \left( R+ 4\partial _ \mu \Phi \partial
^ \mu \Phi- {1 \over
12}(H^1)^2
\right) - {1\over 2} \partial _ \mu \b \partial ^ \mu \b -{1 \over
12} (H^2+ \b H^1)^2
\right]
\eqn\twobstr$$
and the dilaton doesn't couple to the RR fields $\b$, $H^2$ [\bergort]. Note
that there is a consistent truncation to a type I action given by setting $H^1
=0$, $\b=0$.

There are solutions to the field equations derived from \twobstr\  with $\b=0$
and either
$H^1=0$ or $H^2=0$. The action that gives the field equations with
$\b=0$ and   $H^2=0$ is precisely the heterotic string action \het\ with $F=0$
and $H=H^1$, while
the one that gives the field equations with
$\b=0$ and   $H^1=0$ is precisely the type I string action \open\ with $F=0$
and $H=H^2$.
This immediately implies that the type IIB theory has two string and two
five-brane solutions
[\HT,\PKT,\HTE,\HS].
There is a singular NS-NS string and non-singular five-brane given by \strsol,
\fivsol\ respectively with
$H=H^1$, and there is a  RR string given by \opsol\ with $H=H^2$ and a singular
RR
five-brane given by \gensolfiv\ with $H=H^2$ and $\alpha =-1/2$.
The action of the $SL(2,\Z)$ transformation given by \gensolfiv\ with
$\Lambda = \pmatrix { 0 &  1 \cr
-1 & 0} $ interchanges $H^1$ and $H^2$ while $\Phi \to -\Phi$ (with $\b=0$) so
that the two string
solutions and
the two five-brane
solutions are interchanged.

\chapter {Charged Strings,  Zero Modes  and  String-String Duality}

Before considering the zero modes of the ten-dimensional solutions, it will be
useful to first
discuss   the six-dimensional     solutions   [\Seno,\JHS]. The toroidally
compactified six-dimensional heterotic string has an abelian $U(1)^{24}$ gauge
symmetry at generic points in the
$O(4,20)/O(4)\times O(20)$ moduli space.
The fundamental string solution [\DGHR] has two Killing vectors acting without
fixed points
(generating translations in time $t$ and along the string)    so that there is
an  $O(6,22)$ solution generating group: acting on the solution with this group
generates new solutions. A subgroup $O(4,1)\times O(1,20)$ preserves the
boundary conditions and an $O(4)\times O(20)$ subgroup of this acts trivially
[\Sentrans].
Acting with transformations in the $24$-dimensional coset  $O(4,1)/O(4)\times
O(1,20)/O(20)$ generates the charged strings solutions of [\Seno,\Smacrto]
which carry $U(1)^{24}$ electric charge.
They also carry an  electric current proportional to the charge per unit
length, so that these strings are accompanied by both electric  and magnetic
fields.
These solutions are parameterised by the $24$ electric charge densities, and
there are $24$ collective coordinates conjugate to these charges. These
coordinates are periodic, since the charges are quantised, and they are
represented by $24$  world-sheet bosonic coordinates, which are chiral
(right-moving) because the charge and current densities are equal [\Smacrto].
In addition, there are the expected four bosonic zero modes corresponding to
translations in the four dimensions transverse to the string, and four
fermionic ones
corresponding to the unbroken supersymmetries; these are represented by
left-moving world-sheet spinors.
The type II theory compactified on $K_3 \times T^2$ has   a string soliton with
the  same zero mode structure [\Seno,\JHS], which justifies identifying it with
the heterotic string. In particular, the neutral string soliton has charged
string generalisations parameterised by the $24$ electric charge densities,
which equal the current densities, so that the corresponding world-sheet theory
 has $24$ chiral bosons [\Seno].  The type II  origin of these zero modes was
discussed in [\JHS].

The situation is different at the special points in the $O(4,20)/O(4)\times
O(20)$ moduli space at which the gauge symmetry is enhanced to $\hat G=
U(1)^4\times G$ where $G$ is a rank $20$ non-abelian group. The low-energy
effective field theory has a non-abelian gauge symmetry, but there are
solutions in which only gauge fields taking values in a $24$-dimensional Cartan
subalgebra are non-trivial.  The effective theory for such solutions is the
abelian theory considered above, so that there is a $24$ parameter space of
charged string solutions.
As there is such a space for every Cartan subalgebra, there is a charged string
carrying each of the $d_G \equiv dim(\hat G) $ charges. There should be a
chiral bosonic zero mode conjugate to each of these charges, but only $24$ of
them are expected to constitute a mutually commuting set.  Recall that the
chiral sector of the heterotic string can be formulated using the right-moving
sector of a  Wess-Zumino-Witten (WZW)
model  (of level one) instead of  the usual chiral bosons or fermions [\KKhet],
so that here the natural structure to expect
for  the   world-sheet theory describing the extra heterotic zero modes
is  a chiral $\hat G$
WZW
 model (in addition to the usual  bosonic translation zero-modes and the
fermionic modes corresponding to unbroken supersymmetry).
The extra $d_G-24$ zero modes can also be understood in terms of the type II
string.
At the special points in moduli space certain solitonic states  become massless
[\Witten,\HTE], and the soliton has zero-modes in which these extra massless
fields are
excited. The solitonic states that become massless are
 six-dimensional black holes resulting from $p$-brane solitons of the
ten-dimensional theory wrapped around $K_3$ homology cycles, and they become
massless at points in moduli space at which the homology cycles collapse to
zero area [\Witten,\HTE].

The fundamental heterotic string solution of the $SO(32)$ or $E_8\times E_8$
heterotic string and the heterotic soliton of the $SO(32)$ type I string
have a   zero mode structure similar to that of the six-dimensional heterotic
string at enhanced symmetry points.  Solutions
of the effective  $N=1$ supergravity theory coupled to $16$ abelian  vector
multiplets
are solutions of the full theory in which
  only $16$ of the $496$ gauge fields,  taking values in a Cartan subalgebra,
are non-zero.
For any choice of Cartan subalgebra, acting on the fundamental heterotic string
solution \strsol\ with the solution generating transformations in the coset
$O(16,1)/O(16)$ gives charged string solutions [\Smacrto] parameterised by $16$
electric charges.
The collective coordinates  corresponding to these $16$ charges are $16$
periodic chiral bosons.
Repeating this with different choices of Cartan subalgebra, one obtains charged
strings carrying any of the $496$ types of charge, and the corresponding chiral
bosonic zero-modes should be described by a level one WZW model with group
$SO(32)$ or $E_8\times E_8$. These, together with the   eight bosonic zero
modes corresponding to transverse translations and the  eight fermionic zero
modes corresponding to the unbroken supersymmetries give a  zero-mode structure
described by the Green-Schwarz light-cone heterotic string with the heterotic
degrees of freedom described by a chiral WZW model, as in [\KKhet].

Acting on this $496$-dimensional space of charged string generalisations of
\strsol\  with the
field transformation \subs\ (for gauge group $SO(32)$) gives a
$496$-dimensional space of charged string solutions of the type I string action
\open\  generalising the solution \opsol.  These solutions    have the same
singularity structure as \opsol\ and have equal charge and current densities.
The zero mode structure is   that of the $SO(32)$ heterotic string: it is
described by a light-cone string with $8$ bosons, $8$ fermions  and an
$SO(32)$ WZW model, justifying the identification of this soliton with the
heterotic string.

It remains an open question as to whether singular solutions can be acceptable
as solitons and, if so, what types of singularity can be allowed.
The form of a solution of an effective theory can only be trusted down to
length scales corresponding to the masses of the  lightest fields that have
been integrated out, and including such massive fields can drastically change
the short-distance structure of a solution and in some cases remove the
singularity.  Thus even if a solution is  singular, the singularity might be
removed by including extra fields in the effective theory.
In six dimensions, the solitons considered in [\Seno,\JHS] are non-singular.
However, in ten dimensions
 the solution \opsol\ has non-singular space-like sections, but has a
singularity at $r=a$ that can be reached along a finite time-like geodesic, so
that the question arises as to whether it should be regarded as a soliton. It
was argued in [\HTE] that, in the case of the type IIB string, both strings and
both five-branes   should be included in the Bogomolnyi spectrum, despite the
singularity of some of them as solutions of the weakly coupled effective
theory.
Since the uncharged sector of the type I theory is contained within the type
IIB theory, a similar argument implies that the string and five-brane should
both be included in the Bogomolnyi spectrum of the type I string. More
specifically, the arguments of [\HTE] imply that the weakly coupled type I or
IIB string should have a string-like Bogomolnyi state carrying the same charges
as
\opsol\ and so which should be approximately described by the solution \opsol\
at large distances, and have the same zero-mode structure. However, at  short
distances, and in particular near $r=a$, the identification of the Bogomolnyi
state with the classical solution \opsol\ is no longer trustworthy, and other
effects might be expected to come into play.

The singular solution \strsol\ of the heterotic string or type IIB string is
the
solution generated by an elementary string source, and so should not be
regarded as giving rise to
any new states. The  string soliton \opsol\ of the type I string has the   zero
mode structure
of a   heterotic string and in the strong coupling limit becomes an elementary
excitation of the
dual theory, which is the heterotic string theory.
Similarly, the RR string soliton of type IIB is an elementary excitation of the
dual type IIB theory
that emerges at strong  coupling, while the NS-NS string becomes a soliton of
the dual theory.
Indeed, the $SL(2,\Z)$ duality implies that the zero-mode structure of the
RR string soliton should be exactly that required for it to be identified with
the fundamental string of the strongly coupled dual theory.  It is interesting
that in these theories, the strongly coupled
dual theory appears to be obtained by a string/string duality rather than a
string/five-brane duality of the type proposed in [\Strom].


\noindent {\it Note Added:} After this work was completed, a related paper
appeared [\dab] discussing similar issues.

\noindent{\bf Acknowledgements}: I would like to thank Michael Green, Ashoke
Sen and Paul Townsend for valuable
discussions.

\refout
\bye